\documentstyle[prl,aps,psfig,multicol]{revtex}
\newcommand{\be}{\begin{equation}}
\newcommand{\ee}{\end{equation}}
\newcommand{\br}{\begin{eqnarray}}
\newcommand{\er}{\end{eqnarray}}

\def\rg{\rangle}

\begin{document}
\title{Teleportation of a Bose-Einstein condensate state by controlled elastic collisions}
\author{M. C. de Oliveira
}
\address{Departamento de F\'{\i}sica, CCET, Universidade Federal de S\~{a}o
Carlos,\\ Via Washington Luiz km 235, 13565-905, S\~ao Carlos, SP,
Brazil.}
\date{\today}
\maketitle
\draft

\begin{abstract}
A protocol for teleportation of the state of a 
Bose-Einstein condensate trapped in a
three-well potential
 is developed. The protocol uses hard-sphere cross-collision between the condensate modes 
as a means of generating entanglement. As Bell state measurement, it is proposed that a homodyne detection
 of the condensate quadrature is performed through Josephson coupling of the condensate mode
 to another mode in a neighbouring well. 
\end{abstract}
\pacs{03.67.Hk, 03.75.Fi, 32.80.Pj}
\begin{multicols}{2}
\section{Introduction}
Teleportation of quantum states, proposed by Bennett {\it et al.} \cite{bennett}, was first
realised for light polarization states \cite{zeilinger}, owing to the possibility of
 generating nonlocal entanglement between
 parties (the quantum channel) in this system\cite{kwiat}. Although many proposals and experimental
realizations of 
nonlocal
 entanglement of massive particles (atoms and ions) exist \cite{hagley},
 up to now there has been no
experimental evidence of the teleportation of massive particles state \cite{polzik}.
 Indeed, non-linear interactions (a valuable resource for
deterministic generation of entanglement) are always present in
many-particle systems. 
An interesting question then arises - to what extent can the teleportation protocol
 be applied to massive many-particle systems?
A strong candidate
for massive particle state teleportation is the condensate state of a matter field, where non-linear interactions 
appear quite naturally as elastic collisions \cite{walls}.

In this paper we propose an experimental protocol for teleportation 
\cite{bennett} of mode 
states of an atomic Bose-Einstein condensate (BEC) trapped in an optical lattice potential,
 by using controllable elastic collisions  and Josephson coupling \cite{javanainen} 
between modes for both the quantum channel formation and measurement stage. Elastic collisions
are a fundamental
resource for both the formation of the entanglement and the parity operations needed
to correct the teleported state.
 For the measurement stage, we propose a balanced homodyne detection of the BEC modes
quadratures, valid for a small condensate. By measuring the difference of population in 
two condensate modes (the central mode and a reference mode) interacting
via Josephson coupling  the central mode quadrature is determined.

This paper is organized as follows. In Sec. II we present the model for three interacting
 condensate modes trapped in a three-well potential. In Sec. III we propose the teleportation 
protocol using controlled collisions as a means to generate entanglement. In Sec. IV we turn to the measurement stage,
presenting a scheme of homodyne detection of the BEC phase. In Sec. V we present the operations
of parity and displacement needed in order to correct to the required state, the state of the condensate
mode at the receiver station. In Sec. VI we present a physical implementation of the controlled collision on
optical lattices. In Sec. VII a discussion  encloses the paper.
\section{model}
The observation of BECs of diluted trapped neutral atoms 
\cite{1} and the recent achievement of all-optical confinement of a
BEC \cite{alloptical} and condensation on a microelectronic chip \cite{chip} have stimulated a large research program
on BEC of diluted neutral atom gases. Of particular interest is the study of a BEC in a 
confining potential.
In \cite {corney1} the dynamics of a BEC in a double-well potential was modelled.  
In that model,
coherent oscillations due to tunnelling (Josephson-like coupling) \cite{javanainen} between the two wells are suppressed when the number of atoms exceeds
a critical value (self-trapping). 
In fact, the barrier separation between the two wells has a central role as it determines
whether the Josephson coupling between modes is negligible in contrast to cross-collisions,
when the wave-functions of the two modes considerably overlap.
A dynamic process can be envisaged where the two-well barrier is lowered and raised
back adiabatically, such that the elastic collisions leave the two modes in an entangled state \cite{jaksch} - a hallmark of teleportation
protocols. This is the mechanism we focus on here. We  extensively use
controlled collisions between condensate modes, by adiabatically turning off and on the 
potential barriers in an optical lattice potential, in order to teleport the state of one condensate mode, to another mode,
 located inside the trap, but delocalised from the first one. This sequential 
process is depicted in Fig. 1. Initially we
describe the general approach for entanglement generation and measurement and at the end
 we give
a specific but clarifying physical implementation of the time varying potential,
 for optical lattices.
\begin{figure}
\centerline{$\;$\hskip 0 truecm\psfig{figure=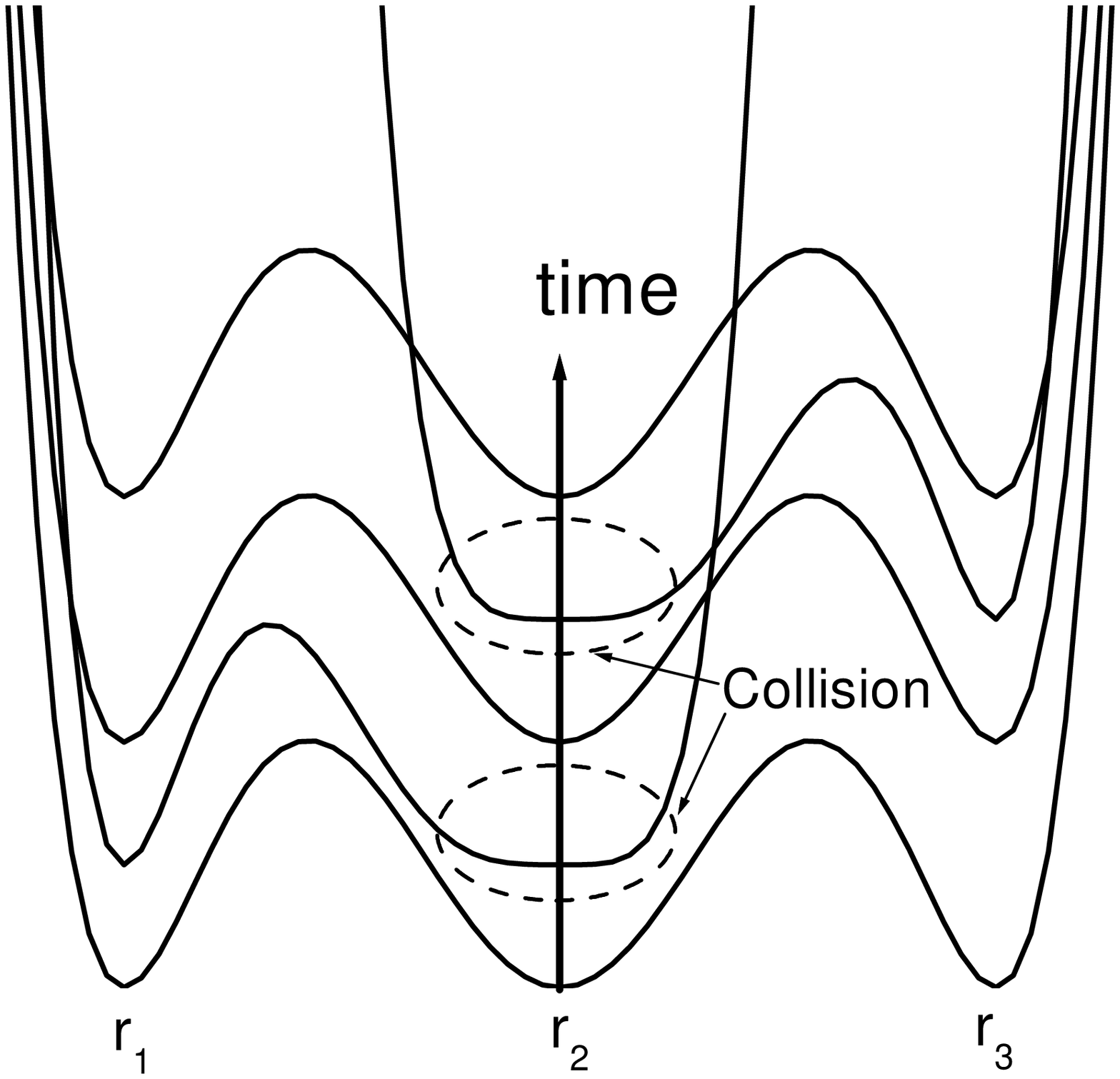,height=6cm,angle=0}}
\vspace{0cm}
\parbox{8cm}{\small Fig.1. Sequential collision for entanglement formation.}
\label{fig1}
\end{figure}

Extending the model in \cite{corney1}, we consider a BEC trapped in a symmetric three-well single-particle
potential $V(r)$ with minima at $r_1$, $r_2$ and $r_3$ disposed along the $z$ axis.
We assume that the three lowest  states of the potential are closely spaced and well
separated from its higher levels, and that many-particle
interactions do not significantly change this situation, to allow a three-mode
 approximation. The potential expanded around each minimum is
\begin{equation}
V({\bf r})=\widetilde V^{(2)}({\bf r}-{\bf r}_j)+...\;\;j=1,2,3,
\end{equation}
where $\widetilde V^{(2)}({\bf r}-{\bf r}_j)$ is the parabolic approximation to the potential
in the vicinity of each minimum. The normalized single-particle ground-state $u_0({\bf r})$ of
the local potential $\widetilde V^{2}({\bf r})$, with energy $E_0$, 
defines the local mode solutions of the individual wells.
 If the position uncertainty in the state $u_0({\bf r})$ is much less than
the separation of the minima of the global potential, the overlap between the modes 
of each well, $\epsilon$, is much less than unity
and the modes are approximately orthogonal\cite{corney1}.
The many-body Hamiltonian describing an atomic BEC in this
potential is 
\begin{eqnarray}
H &=&\int d^{3}x\psi ^{\dagger }({\bf r})\left( -\frac{\hbar }{2m}\nabla
^{2}+V({\bf r})\right) \psi ({\bf r})  \nonumber \\
&&+\frac{1}{2}\frac{4\pi a_{s}\hbar ^{2}}{m}\int d^{3}r\psi ^{\dagger }({\bf %
r})\psi ^{\dagger }({\bf r})\psi ({\bf r})\psi ({\bf r}),  \label{bec}
\end{eqnarray}
where $m$ is the atomic mass, $U_0=4 \pi\hbar^2a/m$ measures the
strength of the two-body interaction, $a$ being the s-wave
scattering length, $\psi({\bf r}, t )$ and $\psi^\dagger({\bf r}, t )$ are the Heisenberg picture
field operators, which annihilate and create atoms at position
${\bf r}$, and normal ordering has been used. In the three-mode
approximation the field operators are expanded in terms of the
local modes and the Heisenberg picture annihilation
and creation operators read as
\begin{equation}
c_j(t)=\int d^3{\bf r} u^*_j({\bf r})\psi({\bf r},t)
\end{equation}
so that $[c_j , c_k]=\delta_{jk}$ to order $\epsilon^0$. 
With this prescription, and retaining terms up to order $\epsilon$, one arrives at two distinct
regimes:
(i) If the potential wells are well separated only self-collision terms are important and the
 many-body Hamiltonian reduces to
\begin{eqnarray}
H_1&=&E_0(c_1^\dagger c_1+c_2^\dagger c_2+c_3^\dagger c_3)\nonumber\\
&&+\hbar\kappa[(c_1^\dagger)^2c_1^2+(c_2^\dagger)^2
c_2^2 +(c_3^\dagger)^2c_3^2],
\end{eqnarray}
where $\kappa=U_0/2\hbar V_{eff}$, and $V_{eff}^{-1}= \int d^3{\bf r}|u_0({\bf r})|^4 $ is the effective mode
volume of each well. In such a situation, no cross-collisions or Josephson tunnelling occur. 
(ii) If the potential wells are not well separated
 Josephson tunnelling  \cite{javanainen} between neighbour wells occurs and
 in the diluted atomic gas regime it prevails over cross-collisions.
The many-body Hamiltonian  then becomes
\begin{eqnarray}
H_2&=&E_0(c_1^\dagger c_1+c_2^\dagger c_2+c_3^\dagger c_3)+
\hbar\frac{\Omega}{2}(c_1^\dagger c_2+c_2^\dagger c_3+ H. c.)\nonumber\\
&&+ \hbar\kappa[(c_1^\dagger)^2c_1^2+(c_2^\dagger)^2
c_2^2 +(c_3^\dagger)^2c_3^2],
\end{eqnarray}
where $\Omega =2{\cal R}/\hbar$ is the tunnelling frequency between two minima, with
\begin{equation}
{\cal R}= \int d^3{\bf r} u_i^*({\bf r})[V({\bf r})-\widetilde V^{(2)}({\bf r}-{\bf r}_i)]u_{i+1}({\bf r}),\;\; i=1,2.
\end{equation}

We assume that in equilibrium state, regime (i) rules out, and the modes can be treated independently.
  If the barrier separating modes 2 and 3 is lowered adiabatically ($\frac{dV}{dt}\ll (E'-E_0)/\hbar$), 
in order to avoid transitions to other states (of energy $E'$),
 the two respective modes overlap and strong cross collision
 occurs as $V_c=2\hbar\kappa c_2^\dagger c_2 c_3^\dagger c_3$. When the barrier is raised back adiabatically,
bringing the system to the equilibrium regime,
the two  modes split again, but now having a non-local entanglement generated
 by the cross collision term, over the time the two modes overlapped.
 In such a case a non-local quantum channel would be formed between condensate modes 2 and 3 
\cite{tel} as we describe in next section. 
\section{Teleportation Protocol}
The following protocol is more efficient if the condensate modes are initially prepared
 in coherent states, but we may assume a general initial state expanding it in the coherent state basis, $|\psi_a\rangle_2=\int d^2\alpha\; a_\alpha\left|\alpha\right\rangle_2$ and
 $|\psi_b\rangle_3=\int d^2\beta\; b_\beta \left|\beta\right\rangle_3$, for modes 2 and 3
respectively. The dynamics governed by $H_1$, together with the cross-collision $V_c$, gives
for modes 2 and 3 alone
\begin{eqnarray}
|\psi(t)\rangle&=&
\int d^2\alpha d^2\beta\;a_\alpha b_\beta
e^{-(|\alpha|^2+|\beta|^2)/2}\nonumber\\
&&\times\sum_{m,n=0}^{\infty}
\frac{1}{\sqrt{m!n!}}
\left(\alpha
e^{-i(E_0/\hbar-\kappa)t}e^{-i\kappa mt}\right)^m \nonumber\\
&&\times \left(\beta e^{-i(E_0/\hbar-\kappa)t}e^{-i\kappa (m+n) t}\right)^n\left|m\right\rangle_{2}\left|n\right\rangle_{3},
\end{eqnarray}
which, for $t=\pi/2\kappa$, turns out to be the entangled state given by
\begin{eqnarray}
\label{a4}
|\Phi(\pi/\kappa)\rangle &=&\frac 1 2\int d^2\alpha d^2\beta\;a_\alpha b_\beta \left[(1-i) |\alpha e^{-i\phi}\rangle_2
|\beta e^{-i\phi} \rangle_{3}\right.\nonumber\\
&&+\left.(1+i)
|-\alpha e^{-i\phi}\rangle_2
|-\beta e^{-i\phi}\rangle_{3}\right]
\end{eqnarray}
where $\phi=(E_0/\hbar-\kappa)/2\kappa$.
Choosing properly the frequency of the modes, $(E_0/\hbar-\kappa)$, a set of
approximately orthonormal states $|\Phi_j\rangle$ can be generated for $E_0/\hbar=(j+1)\kappa$, $j=0,1,2,3$, respectively as
\begin{eqnarray}
\label{bell}
|\Phi_j\rangle&=&\frac 1 2\int d^2\alpha d^2\beta\; a_\alpha b_\beta \left[(1-i)|(-i)^j\alpha\rangle_2
|(-i)^j\beta\rangle_{3}\right.\nonumber\\
&&+\left. (1+i)
|-(-i)^j\alpha \rangle_2 |-(i)^j\beta\rangle_{3}\right].
\end{eqnarray}
From now on we suppose that the condition for the generation of $|\Phi_0\rangle$, {\it i.e.}, $E_0/\hbar=\kappa$, is met. 

A simple teleportation protocol \cite{tel} based on homodyne measurement
 of the condensate modes phases can be performed.
Let us say the condensate mode 1, called hereafter the target mode, is prepared in an
 unknown superposition of the specific form \cite{zoller}
\be
\left|\psi\right\rangle_T=\int d^2\gamma\; c_\gamma\left(A\left|\gamma\right\rangle+B\left|-\gamma\right\rangle\right)
\ee
where $A$ and $B$ are constants respecting normalization conditions. 
Now if the modes
1 and 2 are made to collide the whole condensate state is left 
as
\br\label{final}
&&\frac 1 2\int d^2\alpha d^2\beta d^\gamma a_\alpha b_\beta c_\gamma \;
\left\{-i|\gamma\rg|\alpha\rg(A|\beta\rg-B|-\beta\rg)\right.\nonumber\\
&&+|\gamma\rg|-\alpha\rg(A
|-\beta\rg+B|\beta\rg)+i|-\gamma\rg|\alpha\rg(A|-\beta\rg-B|\beta\rg)\nonumber\\
&&+\left.|-\gamma\rg|-\alpha\rg(A|\beta\rg+B|-\beta\rg)\right\},
\er
a three-partite entangled state composed of four elements. We
 can distinguish each element by the phase of modes 2 and target.
Notice that although the protocol is encoded in continuous variables
states, the protocol itself is discrete as only four equiprobable outcomes are possible. Thus
after the distinction of the target and mode 2 phases, only two bits of classical
 information have to be sent to the mode 3 at the receiver station.
\section{Homodyne detection of BEC phase}
If a joint measurement on both modes 1 and 2 can be envisaged to distinguish the phase 
of each condensate mode,
 the complete Bell state measurement
is realised. 
 Tomographic reconstruction of the
condensate state would allow the distinction between the two different phases. 
Such an approach is based on optical homodyne tomography \cite{bolda},
 using an arrangement composed of an atomic beam splitter and an ideal
atom counter.
Here instead, we describe an alternative scheme for phase determination similar the optical balanced
homodyne measurement, where the Josephson coupling \cite{javanainen} plays the role of an atomic beam splitter.
Consider two condensate modes separated by a barrier, as in \cite{corney1}.
A two-mode approximation is assumed. Neglecting cross-collision terms (once the overlap
 of condensate wave-functions is negligible) the Hamiltonian for the two modes is 
\begin{eqnarray}
H&=&E_0(c^\dagger c+b^\dagger b)+
\frac{\hbar\Omega}{2}(c^\dagger b+b^\dagger c)\nonumber\\
&&+
\hbar \kappa[(c^\dagger)^2c^2+(b^\dagger)^2
b^2],
\end{eqnarray}
Defining new operators as
$S_x=\frac 1 {2N}(c^\dagger c-b^\dagger b)$, $S_y=\frac i {2N}(c^\dagger b-c b^\dagger)$, $
S_z=\frac 1 {2N}(c^\dagger b+c b^\dagger)$,
where $N=\langle c^\dagger c+ b^\dagger b\rangle$,
the equations for the evolution of these operators are
\begin{eqnarray} 
\dot S_x&=&-\Omega S_y,\\
\dot S_y&=&\Omega S_x-2 i \epsilon\Omega S_y-4\epsilon\Omega N S_x S_z,\\
\dot S_z&=&-2 i \epsilon \Omega S_z+4 \epsilon \Omega N S_x S_y,
\end{eqnarray}
where $\epsilon=\frac \kappa \Omega \ll 1$. A semiclassical solution for $S_x$ is given up to 
first-order in $\epsilon$ by
\begin{eqnarray}\label{solu1}
S_x(t)&=&[S_x(0)+\epsilon t (2Nz_0y_0-ix_0)]\cos \Omega t \nonumber\\
&&-[S_y(0)-\epsilon t (2Nz_0x_0+iy_0)]\sin \Omega t, 
\end{eqnarray}
valid only for $\epsilon N\ll 1$, {\it i.e.}, for $\kappa\ll\Omega$, a small number of particles and also for a short time. To derive this 
solution the above operators are expanded as
$S_x=\sum_n \epsilon_n x_n$, $S_y=\sum_n \epsilon_n y_n$ and $S_z=\sum_n \epsilon_n z_n$.
Assuming initially an equal number of atoms in both wells, the solution (\ref{solu1}) simplifies
 to
\begin{eqnarray}\label{homod}
S_x(t)&=&-S_y(0)\sin \Omega t+2\epsilon t N z_0y_0\cos \Omega t +i \epsilon t y_0\sin \Omega t. 
\end{eqnarray}

Consider the following semiclassical picture for the operator $S_y$
\br
\langle S_y\rangle=\frac i 2 |\beta|(\langle c^\dagger\rangle e^{i\theta}-\langle c\rangle e^{-i\theta})
=-{|\beta|}\langle X_{\theta-\pi/2}\rangle,
\er
where the mode B was prepared in a coherent state given by $\beta =|\beta|e^{i\theta}$. It is
easy to observe then that for $\epsilon N\ll 1$, at $t=\pi/2\Omega$, the Eq.(\ref{homod}) gives the well-known result for
balanced homodyne detection, plus a small correction
\be
\langle S_x(\pi/2\Omega)\rangle={|\beta|}\langle X_{\theta-\pi/2}\rangle+i\frac {\pi \epsilon}{2\Omega}  \langle 
y_0\rangle,
\ee
{\it i.e.}, the difference between the numbers of atoms in the two 
wells determines the quadrature phase of one of the matter fields.
This method is, however, sensitive to
the exact determination of the reference phase $\theta$, which for condensates is a
central problem \cite{dunningham}. Here it is simply assumed  that the reference phase 
can be determined by the experimentalist.
With such a quadrature matter field measurement at hand it is possible to distinguish between coherent states like
$|\alpha\rangle$ and $|-\alpha\rangle$, which is the necessary resource to apply to both 
target and mode 2 and thus distinguish
between the many states of the superposition of Eq.(\ref{final}). Remark that the
 requirement for a small number of particles ($N\ll \Omega/\kappa$) avoids the regime of self-trapping,
as observed in \cite{corney1}, when the approximate solution, Eq. (\ref{solu1}), is no longer
valid. This requirement imposes a severe limitation of this detection method to the ``size'' of the BEC to be teleported.
\section{Receiver operations}

The two bits of classical information obtained as described above are transferred to
 the receiver mode 3. 
Now one 
has to apply the operations needed to transform the condensate state mode 3 in the
receiver station into the required state.
Depending on the results of the joint homodyne measurement described above, the
 condensate mode 3 is left in one of the following states,
\begin{eqnarray}\label{unitary}
\label{u1}\int d^2\beta\; b_\beta (A&|\beta\rg+B|-\beta\rg&)\\
\label{u2}\int d^2\beta\; b_\beta (A&|\beta\rg-B|-\beta\rg&)\\
\label{u3}\int d^2\beta\; b_\beta (A&|-\beta\rg+B|\beta\rg&)\\
\label{u4}-\int d^2\beta\; b_\beta(A&|-\beta\rg-B|\beta\rg&)
\end{eqnarray}
In order to transform the condensate states (\ref{u2}), (\ref{u3}), and (\ref{u4}) into the
 required state (\ref{u1}), operations of parity and displacement \cite{tel}, in principle, 
can be realised, as described below.

\subsection{Parity}
A parity operation involving only atomic systems may be envisaged for a two-species BEC 
 with the cross-collision strenght given by $\lambda=U_{AB}/2\hbar V_{eff}$, where $\lambda\neq\kappa=U_{AA}/2\hbar V_{eff}=U_{BB}/2\hbar V_{eff}$ \cite{zoller}.
Supposing the central condensate initially in the state (\ref{u3}) and we require 
to transform it to (\ref{u1}).
 The auxiliary condensate
is prepared in an arbitrary state expanded also in the coherent basis $|\psi\rangle_a=\int d^2\xi a_\xi |\xi\rangle$. Due to cross-collision the state of this system
 at time $t=2\pi/\kappa$ is
\br
|\Phi\rangle_{ac}&=&\int d^2\beta\ d^2\xi; b_\beta a_\xi \sum_n\frac{e^{-\frac 1 2 |\xi|^2}}{\sqrt{n!}}(\xi e^{-i2\pi\frac{(E_0/\hbar-\kappa)}{\kappa}})^n |n\rangle_a\nonumber\\
&&\otimes(A|-\beta e^{-i2\pi\frac{(E_0/\hbar-\kappa+n\lambda)}{\kappa}}\rangle+B|\beta e^{-i2\pi\frac{(E_0/\hbar-\kappa+n\lambda)}{\kappa}}\rangle)_c\nonumber\\
\er
Now suppose a number projective measurement is made on the auxiliary condensate,
 projecting it to $|m\rangle$. The normalized conditioned state of the central condensate is
\br
|\Phi\rangle_c=\frac{_a\langle m|\Phi\rangle_{ac}}{\sqrt{Tr\{|_a\langle m|\Phi\rangle_{ac}|^2\}}}
\er
which for $\lambda=(E_0/\hbar-\kappa)=\kappa/2$ turns out to be
\br
|\Phi\rangle_c&=&\int d^2\beta\; b_\beta (A|\beta e^{im\pi}\rangle+B|-\beta e^{im\pi}\rangle)_c,
\er
and now, if $m$ is even  the central condensate mode is left in the required state,
 if it is not the experiment is to be repeated.
 Thus the number of atoms in the auxiliary condensate mode determines the parity of the central 
condensate mode. In Fig. 2 we plot the efficiency of this proccess by summing over all
 the even number 
probabilities
\be
P_{even}=\sum_{m=even}P(m)=\sum_{m=even}Tr\{|_a\langle m|\Phi\rangle_{ac}|^2\},
\ee
for the auxiliary mode prepared in three distinct states. The dashed line is for the auxiliary 
mode prepared in a number state with average number of atoms $\langle n\rangle$. As it is expected
in this case the probability of success is constantly 0.5 indepently of the atom number.
The dotted line is for a coherent state. We see that when the auxiliary mode is prepared in this
state the probability of success also attains the limit of 0.5, unless the auxiliary mode has less than one atom
in average as it attains the vaccum state. However the state that presents the best efficiency
is the squeezed vacuum state \cite{walls} as depicted by the solid line in Fig. 2, by varying
 the squeezing parameter
$r$ in the top axis of the plot. We see that for $r\le 1.425$ the probability of success of the event is higher than
0.5. This is probably the best situation for parity control.
\vspace{-0.4cm} 
\begin{figure}
\centerline{$\;$\hskip 0 truecm\psfig{figure=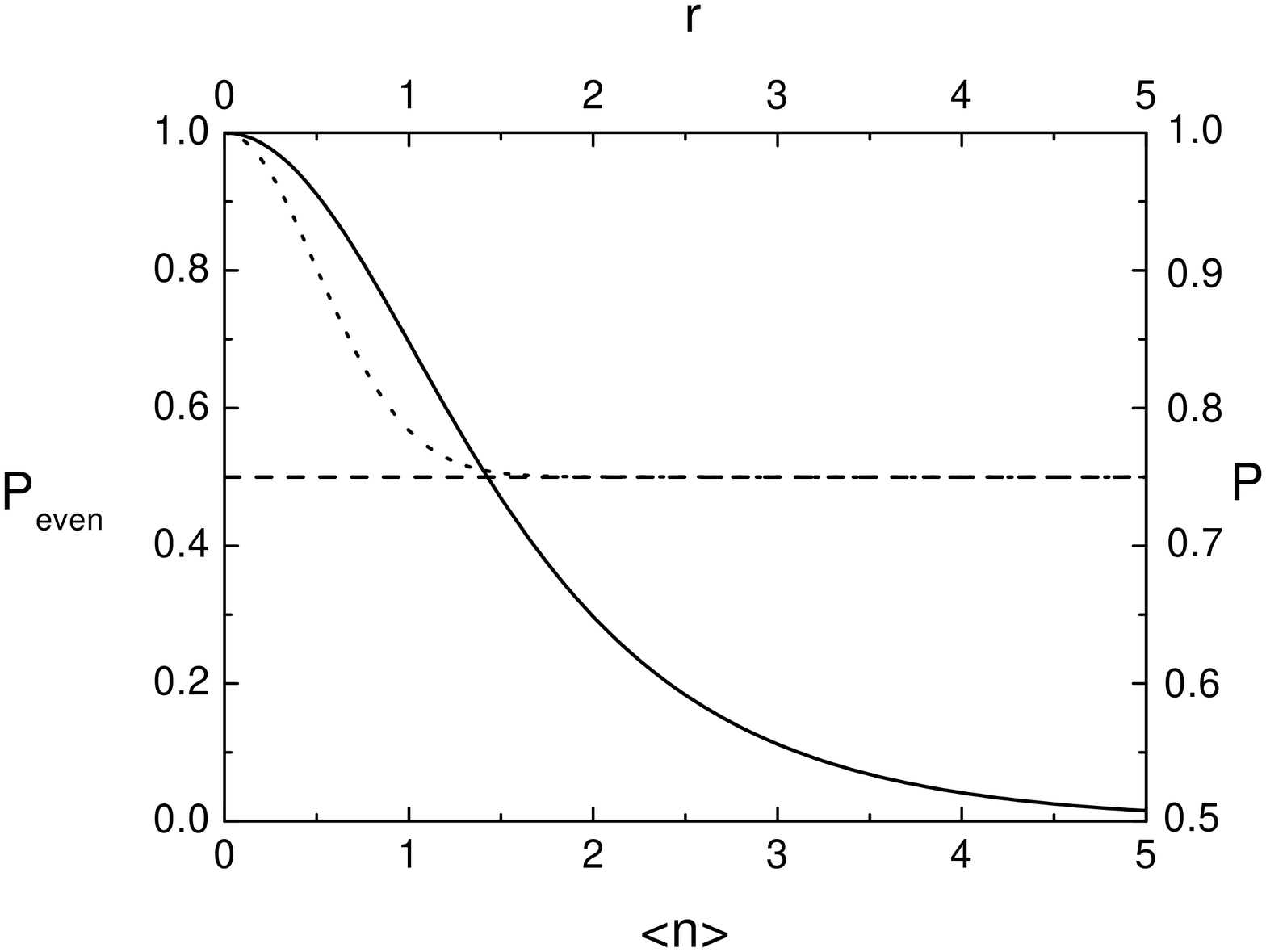,height=7cm,angle=0}}
\vspace{-0.6cm}
\parbox{8cm}{\small Fig.2. Efficiency of even number count event for the auxiliary mode prepared
 in number state (dashed line), coherent state (dotted line), and squeezed vacuum state
 (solid line). Bottom axis represents the number and coherent average number, while the
 top axis is for the squeezing parameter of the squeezed vacuum.
Right axis is for the whole process efficiency for $P_D=1$.}
\label{fig2}
\end{figure}

\subsection{Displacement}
Now, to transform the state (\ref{u2}) into (\ref{u1}), we choose a 
virtual displacement on the central field, defined as follows.
 Consider an atomic beam acting as a
 displacement operator over the central mode
\br
D_\delta|\Phi\rangle_c&=&\int d^2\beta\; b_\beta \left\{\cos[Im(\delta^*\beta)](A|\beta+\delta \rangle-B|-\beta+\delta\rangle)\right.\nonumber\\
&&\left.+i\sin[Im(\delta\beta)](A|\beta +\delta\rangle+B|-\beta+\delta\rangle)\right\}.
\er
Suppose $\delta \in\Re$ and $|\delta|\ll|\beta|$. If $\delta = (l+1/2)\pi/Im(\beta)$,
 for $l=0,1,2...$, the central mode
state is left in the required state (but a global phase of no importance).
The displacement operation can also be directly given by quadrature
($X=b+b^\dagger$) measurement through the homodyne detection described in Sec. IV.
As the parameter $|\delta|\ll|\beta|$ is known to be very small, the
displacement operator, is given by
\be
D_\delta=e^{i|\delta| X}\approx 1+i\delta
X;\;[D_\delta, X]=0.
\ee
Knowing $\delta$, the measurement of $X$
gives the required displacement. The efficiency of this proccess is dependent upon the full knowledge
of the above constants, and thus upon experimental mastering.

Rotations such as that proposed above can be realised by virtual displacement. 
Obviously, the state (\ref{u4}) can be transformed into (\ref{u1}) by sequential applications of the displacement and parity operations.
With this procedure, the teleportation protocol is complete. Notice however that none of the 
above operations are unitary, being dependent on selective measurements, and thus irreversible.
A figure of merit of the whole proccess can be given by adding the probability of succes of 
each event and dividing by the number of equiprobable events,
\be
P=\left(1+P_{even}+P_D+P_{even}P_D\right)/4,
\ee
where $P_D$, the probability of success of the displacement operation, is determined by the 
experimental control. 
In the right axis of Fig. 2 we compare $P$ for the three states considered above for 
the auxiliary mode, fixing $P_D=1$. Again we observe that for $r\le 1.425$ the squeezed vacuum is the best state
for the auxiliary mode to be prepared. In Fig. 3 we analyse $P$ for the squeezed vacuum 
by varying both $r$ and $P_D$. The efficiency decreases considerably (less than $0.3$) 
when both, the squeezing parameter is high and the displacement process efficiency is low.
 But, when
the squeezing parameter is fixed to 0, or alternatively $P_D$ is fixed to 1, either situations have efficiency
higher than $0.5$ attaining the ideality for $P_D=1$ and $r=0$, the vacuum state. This 
situation corresponds to an empty mode. Thus if the auxiliary mode can be initially prepared
in a vacuum state the protocol has high probability of success.

\begin{figure}
\centerline{$\;$\hskip 0 truecm\psfig{figure=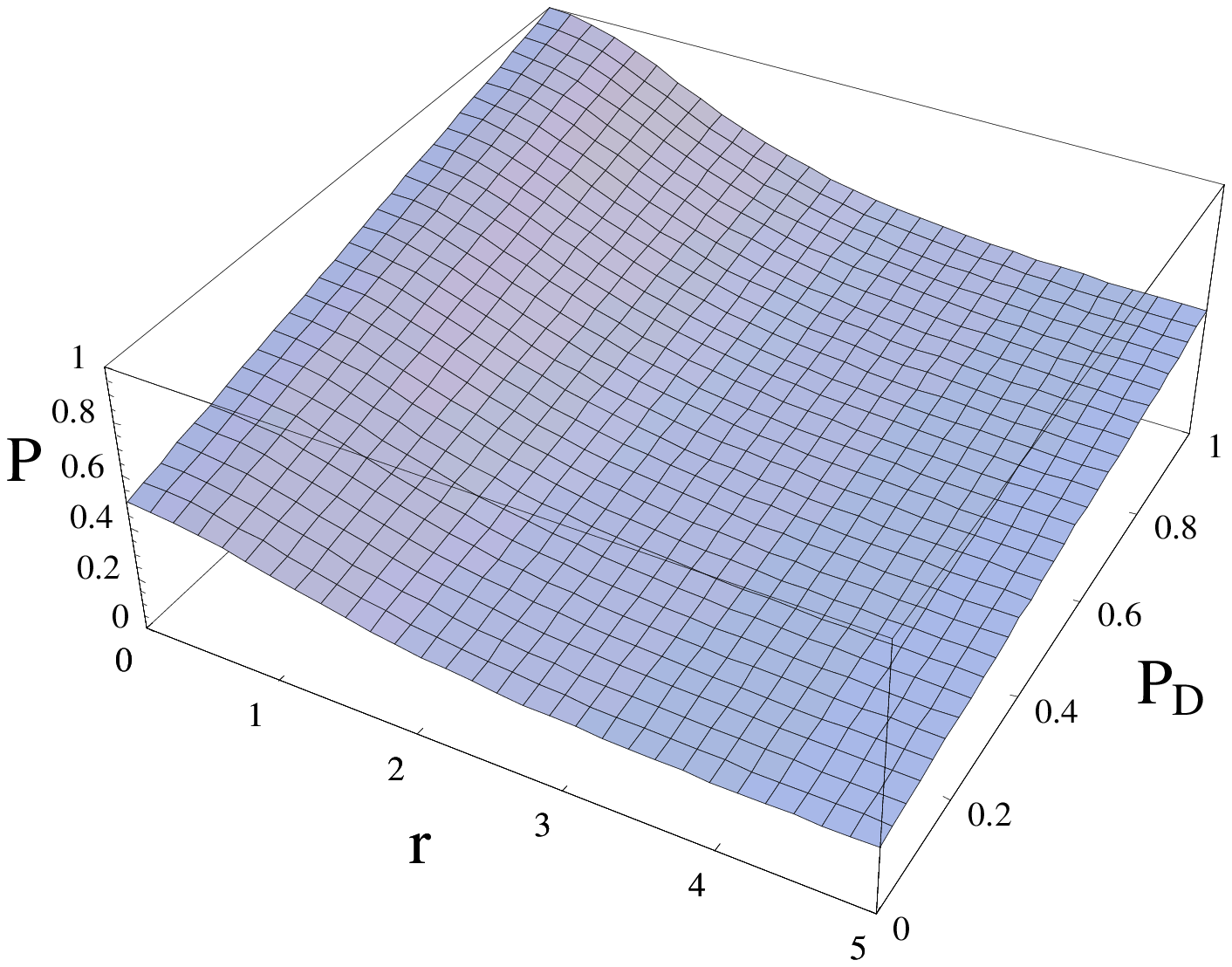,height=14cm,angle=0}}
\vspace{-6.0cm}
\parbox{8cm}{\small Fig.3. Efficiency of success of operation for
 the auxiliary mode prepared in a squeezed vacuum state.}
\label{fig3}
\end{figure}

\section{physical implementation}
A specific physical implementation of the
time dependent potential can be designed, similarly to what is presented in  \cite{deutsch} on optical 
lattices. Let us consider a picture of an atom driven on a $|J=1/2\rangle\rightarrow|J=3/2\rangle$ transition by
a one-dimensional optical lattice red-detuned in 1D lin-angle-lin configuration. The optical field
can be written as a superposition of opposite helicity standing waves \cite{deutsch,finkelstein}
\br
{\bf E}_L(z)&=&\sqrt2 E_1 [-e^{-i\theta/2}\cos(k_Lz+\theta/2){\bf e}_+\nonumber\\
&&+e^{i\theta/2}\cos(k_Lz-\theta/2){\bf e}_-],
\er
for a convenient choice of relative phase between the beams. The potential for atoms in the ground state
is
\br
U(z)&=&-\frac{2U_1}{3}\{2[1+\cos\theta\cos(2k_Lz)]I\nonumber\\
&&+[\sin\theta\sin(2k_Lz)]\sigma_z\}-\frac\hbar2\gamma{\bf B}\cdot{\bf \sigma},
\er
where $U_1$ is the light shift produced by a single beam of amplitude $E_1$, while ${\bf B}$ is a
magnetic field and $\{I,\sigma_i\}$ are the identity and
 Pauli spin operators in the ground-state manifold. By varying $\theta$ the peak-peak modulation depth of the potential and the distance between the $|m=1/2\rangle$ and $|m=-1/2\rangle$ potential wells
are changed by
\be 
U_p=\frac 4 3 U_1 \sqrt{3\cos^2\theta+1},\;\;\;k_l\Delta z=\tan^{-1}\left(\frac{\tan\theta}2\right),
\ee
respectively, while changing the longitudinal component of ${\bf B}$
 shifts the minima of these wells. The transverse component of ${\bf B}$ breaks the degeneracy
 of the bipotential at positions of linearly polarized light. By 
varying $\theta$ appropriately one can design adiabatically time dependent 
potentials \cite{jaksch,deutsch}. By loading the optical potential with the condensate at $\theta=\pi/2$, and 
changing it from $\pi/2$ to $5\pi/2$ adiabatically, we obtain the sequential controlled collision
 we required, as pictured in Fig. 4. The three condensate modes should be loaded in neighbouring wells in order to produce
the required protocol. 

\begin{figure}
\centerline{$\;$\hskip 0 truecm\psfig{figure=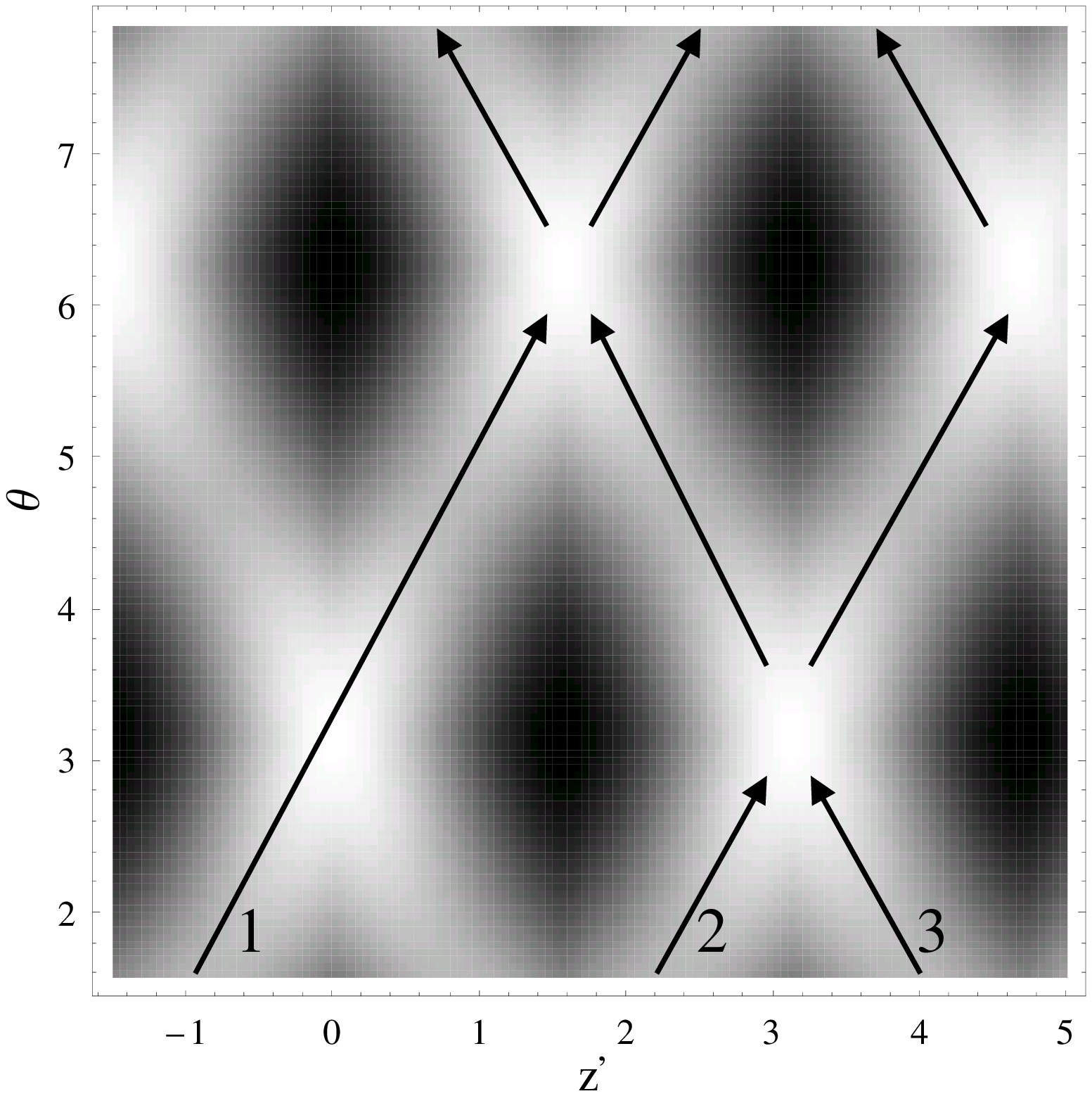,height=12cm,angle=0}}
\vspace{-3.5cm}
\parbox{8cm}{\small Fig.4. Density map of the $m=1/2$ optical lattice trap potential.
$B_{\Vert}=0$, while $B_\bot\neq 0$. 
 Brighter zones corresponds to potential minima, $z^\prime=2k_L z$.
The sequential collision path is depicted by the numbered arrows.}
\label{fig2}
\end{figure}

 An alternative approach can be designed much in the same way of
 the controlled 
 entanglement generation in atomic ensembles of Jaksch {\it et al.} \cite{jaksch},
with no need of a magnetic field, but where different internal spin components are selected
to promote the minima of the potential to move in relation to each other. Adiabaticity here,
 is then related to the speed of the the minima displacement. This alternative has an 
advantage in relation to the first one, 
which is that the condensate modes do not need to be loaded in neighbouring wells, 
as longer as the condensate modes 1 and 3 have the same
spin component, the mode 2 with an opposite component, can be moved in relation to the other
 two modes effecting the required 
operations, despite their (mode 1 and 3) relative position.
\section{Discussion}

In summary, we have speculated about teleportation of a BEC state between modes in a
three-well potential (over short distances). Hard-sphere cross-collision is used 
as a resource to generate entanglement between modes. For the final state measurement, the
 protocol employs
a homodyne detection of the 
BEC quadrature, in which
Josephson coupling of the condensate mode to an auxiliary mode in a neighbouring
 well plays the role of atomic beam-splitter. 

The teleportation protocol itself was constituted of three stages, (i) entanglement
formation; (ii) measurement; and (iii) receiver operations. In (i) we limited ourselves to the 
situation of adiabatic time varying potential ($\frac{dV}{dt}\ll (E'-E_0)/\hbar$) to avoid excitations
to other states. In (ii) the homodyne detection of the 
BEC quadrature guarantees that  for ($N\ll \Omega/\kappa$), a well-diluted atomic gas and short time
of interaction, a condensate mode quadrature can be determined by the difference of atoms between the
central mode and an auxiliary mode. In (iii) parity and displacement operations were proposed to fix the teleported state
at the receiver station. 
Since those operations are dependent upon selective measurements, the whole teleportation
 protocol is limited by the operations efficiencies. The efficiency of the parity operation is given by
the even-parity of number of atoms present in
 the auxiliary state, which can be higher than 0.5 once the auxiliary condensate can
 be prepared approximately to a vacuum squeezed state with squeezing parameter $r\le 1.425$. Together
with the virtual displacement operations, 
those parity operations have their efficiency limited by the full knowledge of the experimental parameters. 
Such is the case for the squeezing parameter, $r$, and the virtual displacement $\delta = (l+1/2)\pi/Im(\beta)$,
 for $l=0,1,2...$. The efficiency of the whole operation vary from 0.3 to 1 for auxiliary mode prepared
in the squeezed vacuum, by varying $r$ and the probability of success of the displacement operation.

We stress that the proposed protocol
is idealised in that the above measurement processes will be very sensitive to the 
presence of dissipation, such as collisions with non-condensate atoms \cite{anglin}.
We expect that, together with the requirement of a small number of particles at the measurement
 stage ($N\ll \Omega/\kappa$), decoherence effects impose a severe limitation to the ``size" of the BEC to be 
teleported. 
 The competition between Josephson coupling and cross-collision could also represent 
a limitation
for the above protocol. We know that if the modes are prepared in coherent states
this does not represent any problem, once the Josephson coupling just adds a phase to the
 state. However for other states this may not be so. Thus the competition of those evolutions need
to be further investigated.

As a last remark, throughout the paper we have assumed a coherent basis 
representation for the condensate mode states, which are better adapted to our discussion. Although
not specifically adressing to coherent states for the modes, those states could also be considered, if each mode
were actually entangled with another condensate to form an
entangled coherent state. The scheme could then proceed as described and
would effectively be entanglement swapping. 

Despite the idealisation, we hope that the above considerations bring some contribution
to the realisation of matter field state teleportation in the near future.

\acknowledgments

{The author thanks G. J. Milburn for encouraging comments, and M. H. Y. Moussa and G. A. Prataviera for 
enlightening discussions. This work is supported by FAPESP-Brazil,
 under projects 01/00530-2 and 00/15084-5.
}
%

\end{multicols}
\newpage

\begin{references}
\bibitem{bennett}
C.H. Bennett, G. Brassard, C. Crepeau, R. Jozsa, A. Peres, and
W.K. Wootters, Phys. Rev. Lett. {\bf 70}, 1895 (1993).
\bibitem{zeilinger}
D. Bouwmeester J.-W. Pan, K. Mattle, M. Eibl, H.
Weinfurter, and
A. Zeilinger, Nature(London) {\bf390}, 575 (1997).
\bibitem{kwiat}
P. G. Kwiat, K. Mattle, H. Weinfurter, A. Zeilinger, 
A. V. Sergienko, and Y. Shih, 
Phys. Rev. Lett. {\bf 75}, 4337 (1995). 
\bibitem{hagley}E. Hagley, X. Ma\^\i tre, G. Nogues, C. Wunderlich, M. Brune, J. M. Raimond, 
and S. Haroche, Phys. Rev. Lett. {\bf 79}, 1 (1997);
A. Rauschenbeutel, G. Nogues, S. Osnaghi, P. Bertet, M. Brune,
J.M. Raimond, and S. Haroche, Science {\bf 288}, 2024 (2000);
C.A. Sackett, D. Kielpinski, B.E. King, C. Langer, V. Meyer, C.J. Myatt,
M. Rowe, Q.A. Turchette, W.M. Itano, D.J. Wineland, and C. Monroe, Nature {\bf 404}, 256 (2000);
 B. Julsgaard, A. Koxhekin, and E.S. Polzik, Nature {\bf 413}, 400 (2001).
\bibitem{polzik}A. Kuzmich and E.S. Polzik, Phys. Rev. Lett. {\bf 85}, 5639 (2000); 
Lu-Ming Duan, J. I. Cirac, P. Zoller, and E. S. Polzik, {\it ibid.}{\bf 85}, 5643 (2000). 
\bibitem{walls}A.S. Parkins and D.F. Walls, Phys. Rep. {\bf 303}, 1 (1998).
\bibitem{javanainen}J. Javanainen, Phys. Rev. Lett. {\bf 57}, 3164 (1986);
 A. Smerzi, S. Fantoni, S. Giovanazzi, and S. R. Shenoy, {\it ibid.} {\bf 79}, 4950 (1997).
\bibitem{1}M.H. Anderson, J.R. Ensher, M.R. Matthews, C.E. Wieman, and E.A. Cornell,
Science {\bf269}, 198 (1995);
K.B. Davis et al., Phys. Rev. Lett. {\bf75}, 3969 (1995).
\bibitem{alloptical}M. Barrett, J. Sauer, and M.S. Chapman, Phys. Rev. Lett. {\bf87},
 010404 (2001).
\bibitem{chip}W. H\"ansel, P. Hommelhoff, T.W. H\"ansch, and J. Reichel, Nature {\bf 413}, 498 (2001).
\bibitem{corney1}G.J. Milburn, J. Corney, E.M. Wright, and D.F. Walls, Phys. Rev. A {\bf55}, 4318 (1997);
J. F. Corney and G.J. Milburn, Phys. Rev. A {\bf 58}, 2399 (1998).
\bibitem{jaksch}D. Jaksch {\it et al.}, Phys. Rev. Lett. {\bf 82}, 1975 (1999). 
\bibitem{tel}M.C. de Oliveira and G.J. Milburn, Phys. Rev. A {\bf65}, 032304 (2002).
\bibitem{zoller}J.I. Cirac, M. Lewenstein, K. M\o lmer, and P. Zoller, Phys. Rev. A {\bf 57}, 1208 (1998);
D. Gordon, and C.M. Savage, {\it ibid.} {\bf59}, 4623 (1999).
\bibitem{bolda}E.L. Bolda, S.M. Tan, and D.F. Walls, Phys. Rev. Lett. {\bf79}, 4719 (1997);
R. Walser, {\it ibid.} {\bf 79}, 4724 (1997); 
S. Mancini and P. Tombesi, Europhys. Lett. {\bf 40}, (1997).
\bibitem{dunningham}J. A. Dunningham and K. Burnett, Phys. Rev. Lett. {\bf 82}, 3729 (1999).
\bibitem{deutsch} I.H. Deutsch and P.S. Jessen, Phys. Rev. A {\bf 57}, 1972 (1998).
\bibitem{finkelstein} V. Finkelstein, P.R. Berman, and J. Guo, Phys. Rev. A {\bf 45}, 1829 (1992).
\bibitem{anglin}J. Anglin, Phys. Rev. Lett. {\bf 79}, 6 (1997).
\end{references}
\end{document}